\documentclass[preprint,aps,floats,nofootinbib,amssymb]{revtex4}
\pdfoutput=1 
\usepackage[T1]{fontenc} 
\usepackage{hyperref}
\usepackage[utf8]{inputenc}
\usepackage{amssymb}
\usepackage{latexsym}
\usepackage{graphics}
\usepackage{graphicx}
\usepackage{feynmf}
\usepackage{color}
\usepackage[sort&compress]{natbib}
\usepackage{epsf,epsfig}
\usepackage{amsmath}
\usepackage{hyperref}
\usepackage{subfigure}
\usepackage{mathrsfs}

\newcommand{\be}{\begin{equation*}}
\newcommand{\ee}{\end{equation*}}
\newcommand{\ba}{\begin{eqnarray*}}
\newcommand{\ea}{\end{eqnarray*}}

\newcommand{\bw}{\begin{widetext}}
\newcommand{\ew}{\end{widetext}}

\begin{document}

\title{\vspace*{1.in} \large
Bouncing neutrons, Equivalence Principle and Unruh-like radiation}
\vspace*{0.5cm}

\author{Dom\`enec Espriu\footnote{espriu@icc.ub.edu}}
\affiliation{
Departament de F\'isica Qu\`antica i Astrof\'isica and
Institut de Ci\`encies del Cosmos (ICCUB), \\
Universitat de Barcelona, 
Mart\'i Franqu\`es 1, 08028 Barcelona, Spain}

\begin{abstract}
  We take a fresh look at the so-called bouncing neutron experiments, and we
  study the quantum dynamics of ultra-cold neutrons above a reflecting horizontal mirror
  from the point of view of the  Equivalence Principle of General Relativity.
  The starting point is the gravitationally bound-state problem: the neutron occupies discrete Airy-function
  states in the Earth's gravitational field, and a periodic vertical displacement of the mirror
  induces transitions between them. We formulate this effect in a non-inertial frame, using the
  equivalence principle to represent the mirror motion as a time-dependent fictitious gravitational
  potential. This allows us to quantitatively explain some features of the neutron bouncing experiments.
  We then ask a logically subsequent question: the same prescribed boundary motion, treated at the level of quantum
  field theory in the non-inertial frame, can generate electromagnetic radiation through the dynamics of the
  field in the accelerated frame. This radiation is non-thermal and is therefore distinct from the thermal
  spectrum associated with the standard Unruh effect. Since the neutron is electrically neutral but carries
  a magnetic dipole moment, the resulting electromagnetic field may nevertheless couple to its spin.
  We estimate the acceleration required for this effect to become measurable; for neutrons
  this lies far beyond laboratory capabilities, but the scaling laws we derive locate the
  threshold and suggest that electrons in laser wakefields may be more suitable probes.
\end{abstract}

\maketitle

\section{Introduction}
The quantum motion of ultra-cold neutrons (UCN) above a horizontal reflecting surface provides a particularly
clean realization of gravitational bound states. This experimental line was initiated by
Nesvizhevsky and collaborators\cite{nature} who convincingly showed the quantization of the Earth's classical
gravitational potential after the pioneering idea of \cite{pioneer}.
The vertical spectrum is determined by the Earth's gravitational field and the boundary conditions,
leading to peV energy splittings and wave functions extending over micrometre scales \cite{details}.
Experimentally it was well established that a controlled
oscillation of the mirror provides a periodic perturbation that can drive transitions between these states
and the first quantum levels were revealed. These experiments are termed q-BOUNCE in the literature \cite{name}.

The one-dimensional hamiltonian describing this physical system is
\begin{equation}
H= \frac{p^2}{2 m_i} + m_g g z.
\end{equation}
In this formula a distinction has been made between the inertial and gravitational masses. 
By measuring the quantum levels in the Earth's gravitational field, q-BOUNCE experiments set bounds
on the validity of Weak Equivalence Principle (WEP) $ m_i=m_g $ \cite{WEP}. The results depend on the
ratio of inertial and gravitational masses, yielding
\begin{equation}
  \frac{\delta m}{m} < 1.9 \times 10^{-6}.
\end{equation}
These experiments have also been able to set limits on non-Newtonian fifth-force-like interactions
between the neutron and the Earth, parameterized by a Yukawa-type
potential $V_{\text{Yuk}}(z) = -\alpha G \frac{M_{\oplus} m}{z} e^{-z/\lambda}$.
A quoted limit at micrometer length scales ($\lambda \sim 10\,\mu\mathrm{m}$) is \cite{5th}
\begin{equation}
  |\alpha| < 10^8 \quad \text{at } \lambda = 10\,\mu\mathrm{m}.
\end{equation}

While the evidence for neutrons filling in the quantized levels of the Earth's potential
seems now well established, the q-BOUNCE experiments do not provide an intuitive rationale for some of the values
that they adopt in the experimental settings. For instance, for the $z=0$ surface they use a
beryllium mirror oscillating at
mechanical frequencies adjusted to the expected theoretical level separations and amplitudes around $A=1 \mu$m.
These values are determined by optimization so as to maximally populate the various levels. Cold neutrons enter the experiment
with a velocity of about 7 m/s and they leave the experimental setup in less than a second. Can it be understood
that cold neutrons settle so quickly in the lowest levels of the Earth's gravitational field?

In this work we will see that the Equivalence Principle allows us to gain a simpler understanding
of the experiment and justify some of the assumptions and settings. The LAB reference frame
is inertial, but the neutrons that are subject to a Dirichlet boundary condition at $z=0$
(the mirror surface) are in fact in a non-inertial frame. According to General Relativity,
this is fully equivalent to a fictitious gravitational field that adds up to the Earth's potential.
If one places itself in this non-inertial frame there is no mirror, just gravity. This simple fact
is sufficient to account for many of the details of the experiment. For instance, if a neutron
decays to a lower level or is excited to a higher level of the external potential, a (fictitious)
gravity quantum is emitted or absorbed.

A further question arises when the electromagnetic field is quantized in the non-inertial description.
The time-dependent transformation associated with the mirror motion affects the field modes and, as we shall see,
leads to particle production. This is the analog of the Unruh effect\cite{unruh} in the
present context. In particular, photons will be generated, carrying electromagnetic fields that
could in principle interact with the neutron magnetic moment and induce spin transitions.
The resulting effect is expected to be exceedingly small. In this work we quantify it for the mechanical mirror
settings employed in q-BOUNCE experiments and also consider magnetically induced oscillations,
thereby assessing the feasibility of observing radiation associated with this non-inertial setting.

All these phenomena could have an interpretation in the inertial frame too, but it is much less
obvious in some cases. For instance, a moving boundary can parametrically mix electromagnetic field modes and,
under appropriate conditions, generate real photons through the so-called dynamical Casimir
effect (DCE) \cite{moore,DCE,chalmers}. Those photons carry an oscillating magnetic field which can couple to the neutron
magnetic dipole moment. Yet, the Equivalence principle allows for a clean and simple derivation.

This work is organized as follows: the gravitational quantum states and
their mechanically induced transitions are established first;
a full quantum interpretation regarding the fictitious gravitational field is introduced next. Then
the electromagnetic vacuum-radiation resulting from the non-inertial description 
is introduced and possible experimental evidence discussed. As mentioned, the treatment is purely
field theoretical using the tools of General Relativity, avoiding references to the lab inertial frame.

\section{Gravitational quantum bound states}
\label{sec:boundstates}
We consider a non-relativistic neutron of mass $m=m_i=m_g$ above a perfectly reflecting horizontal mirror at $z=0$ in
the Earth's gravitational field. The effective optical potential governing ultra-cold
neutron reflection at material boundaries is 
standardly treated in neutron optics (see, e.g., \cite{ignatovich}).

The vertical wave function satisfies
\begin{equation}
    -\frac{\hbar^2}{2m}\frac{d^2\psi_n(z)}{d z^2}+mgz\,\psi_n(z)=E_n\psi_n(z),
\end{equation}
with the boundary condition $ \psi_n(0)=0$.
Let us introduce some natural length and energy scales \cite{details}
\begin{equation}
    z_0=\left(\frac{\hbar^2}{2m^2g}\right)^{1/3},
    \qquad
    E_0=\left(\frac{m\hbar^2g^2}{2}\right)^{1/3}=mgz_0,
\end{equation}
the normalized stationary states may be written as
\begin{equation}
    \psi_n(z)=C_n{\rm Ai} \left(\frac{z}{z_0}-\alpha_n\right),
\end{equation}
where $\alpha_n>0$ are defined by ${\rm Ai}(-\alpha_n)=0$,
and ${\rm Ai}(z)$ is the Airy function.

The normalization constant is
\begin{equation}
    C_n=\frac{1}{\sqrt{z_0}\,|{\rm Ai}'(-\alpha_n)|}.
\end{equation}

The discrete energy spectrum is
\begin{equation}
    E_n = E_0\alpha_n = \hbar \omega_0 \alpha_n,
    \qquad
    \omega_0 \equiv \frac{E_0}{\hbar} = \frac{m g z_0}{\hbar} \simeq 914.1\,\mathrm{rad/s}.
\end{equation}
For the two lowest states,
\begin{align}
    \alpha_1&\simeq 2.33811, & E_1&\simeq 1.407\,\mathrm{peV},\\
    \alpha_2&\simeq 4.08795, & E_2&\simeq 2.461\,\mathrm{peV}.
\end{align}
More generally, any transition between two gravitational levels $n$ and $k$ ($n \neq k$) has
frequency $\omega_{nk} \equiv (\alpha_k - \alpha_n)\omega_0$ and exact Airy dipole matrix element \cite{details}
\begin{equation}\label{airy_dipole}
    |\langle k|z|n\rangle| = \frac{2z_0}{(\alpha_k - \alpha_n)^2}.
\end{equation}
For the fundamental $1 \leftrightarrow 2$ transition, this gives $\Delta E_{12} = E_2 - E_1 \simeq 1.054\,\mathrm{peV}$
and $ \omega_{12} = \Delta E_{12}/\hbar$.

The vertical extent of the states is macroscopic on atomic scales. In particular,
\begin{equation}
    \langle 1|z|1\rangle\simeq 9.15\,\mu\mathrm{m}, \qquad
    \langle 2|z|2\rangle\simeq 16.00\,\mu\mathrm{m}, \quad\ldots \quad
     \langle n|z|n\rangle=\frac{2}{3}\alpha_n z_0.
 \end{equation}

\subsection{Two-level description}
For a controlled perturbation near the $1\leftrightarrow2$ resonance, it is useful to restrict the neutron
Hilbert space to these two states. Writing the off-diagonal perturbation as
\begin{equation}\label{rabi2level}
    H_{\mathrm{drive}}(t)=\lambda\cos(\Omega t)\left(|2\rangle\langle1|+|1\rangle\langle2|\right),
\end{equation}
one obtains
\begin{equation}
    P_{1\to2}(t)=\sin^2\left(\frac{\lambda t}{2\hbar}\right),
\end{equation}
when $\Omega=\omega_{12}$. Away from exact resonance we use the generalized Rabi frequency
\begin{equation}
    \Omega_R=\sqrt{(\omega_{12}-\Omega)^2+\frac{\lambda^2}{\hbar^2}},
\end{equation}
and the corresponding two-level transition probability is
\begin{equation}
    P_{1\to2}(t)=\frac{\lambda^2}{\hbar^2}
    \frac{\sin^2\left(\Omega_R t/2\right)}{(\Omega_R)^2}.
\end{equation}

\section{The oscillating mirror and the non-inertial frame}
\label{sec:noninertial}
The reflecting surface can be driven harmonically. We choose throughout the paper the convention
\begin{equation}\label{phase}
    z=z_{\mathrm{LAB}}-A\cos(\Omega t_{\mathrm{LAB}}),
\end{equation}
where $A$ is the displacement amplitude and $\Omega$ is the angular frequency. The mirror is located
at $z=0$. Its acceleration is
\begin{equation}
    a(t_{\mathrm{LAB}})=-A\Omega^2\cos(\Omega t_{\mathrm{LAB}}).
\end{equation}

In the frame comoving with the mirror the boundary is stationary, while an inertial (fictitious) force appears.
To leading order this force is represented by the potential
\begin{equation}
    V_{\mathrm{in}}(z,t)=-mA\Omega^2 z\cos(\Omega t).
\end{equation}
Locally, the acceleration of the frame can be described by a time-dependent gravitational field with
effective acceleration opposite to the frame acceleration.

At linear order in the displacement amplitude, the fictitious gravitational potential is described by the metric
\begin{equation}
    g_{00}=c^2(1+h_{00}),
    \qquad
    h_{00}\simeq -\frac{2A\Omega^2 z}{c^2}\cos(\Omega t),
\end{equation}
There is an associated time transformation too
\begin{equation}\label{timechange}
  t= t_{\mathrm LAB} + \frac{A \Omega}{c^2} z_{\mathrm LAB} \sin \Omega t_{\mathrm LAB} .
\end{equation}
This transformation keeps the transformed metric diagonal, and physically it means clocks at different
spatial points are synchronized by light signals (Einstein synchronization). The time $t$ at $z=0$ does not
exactly correspond to the proper time of the neutron, but it differs from it at $O(A^2)$ only. We claim
accuracy of our results at $O(A)$ only.

The transformation also generates a $h_{zz}$ component
\begin{equation}
h_{zz}= - \frac{A^2 \Omega^2}{c^2} \sin^2 \Omega t ,
\end{equation}
but it would couple to a very small component of the neutron's energy-momentum tensor
In addition, it is $O(A^2)$.

General relativity dictates the matter-gravity interaction
\begin{equation}
    H_{\mathrm{int}}(t)=\frac12\int d^3x\,T^{\mu\nu}h_{\mu\nu}.
\end{equation}
In the non-relativistic limit all components of the neutron energy-momentum tensor can be neglected except 
\begin{equation}
    T^{00}(\mathbf{x})\simeq mc^2\delta^3(\mathbf{x}-\mathbf{z}),
\end{equation}
so that the leading interaction reduces to the same inertial potential,
\begin{equation}
    H_{\mathrm{int}}(t)=-mA\Omega^2 z\cos(\Omega t).
\end{equation}

This equality is a consequence of the Equivalence Principle: the same measurable transition can be described
either by an inertial force in the comoving frame or by the corresponding weak-field metric perturbation
in General Relativity which fixes in a well prescribed manner the coupling of $h_{\mu\nu}$ to matter.

\section{Transition matrix element}
\label{sec:transition}
The off-diagonal matrix element relevant for the fundamental transition between the two lowest levels is
\begin{equation}
    \langle2|H_{\mathrm{int}}(t)|1\rangle
    =-mA\Omega^2\cos(\Omega t)\,\langle2|z|1\rangle.
\end{equation}
For the Airy states one finds \cite{details}
\begin{equation}
    \langle2|z|1\rangle
    =\frac{2z_0}{(\alpha_2-\alpha_1)^2},
\end{equation}
implying that the coupling parameter of the two-level Hamiltonian Eq. (\ref{rabi2level}) is
\begin{equation}
    \lambda=mA\Omega^2\frac{2z_0}{(\alpha_2-\alpha_1)^2}
    =\frac{2A\Omega^2}{(\alpha_2-\alpha_1)^2}
      \frac{\hbar^{2/3}m^{1/3}}{2^{1/3}g^{1/3}}.
\end{equation}
Numerically,
\begin{equation}
    \frac{2}{(\alpha_2-\alpha_1)^2}\simeq 0.653,
\end{equation}
so  the gravitational-state Rabi frequency at resonance is
\begin{equation}
    \Omega_R=\frac{\lambda}{\hbar}
    =A\Omega^2\,0.653\left(\frac{m}{2\hbar g}\right)^{1/3}.
\end{equation}

Let us now make contact with the experimental parameters of the q-BOUNCE experiments.
First of all, as expected, the frequency of the mechanical vibration has to coincide with
the one associated to $\omega_{12}$. Experiments \cite{nature} use this value
that keeps the neutron population oscillating between the first two levels.

As reported in \cite{nature}, the interaction time is approximately
\begin{equation}
    \tau=\frac{L}{v_x}\simeq 0.026\,\mathrm{s}
\end{equation}
resulting from neutrons crossing the mirror section at $v_x\simeq 7$ m/s.
A formal $\pi$-pulse condition at exact resonance ($\Omega_R \tau=\pi$) yields a required displacement amplitude 
of $A \simeq 772\,\mathrm{nm}$ (or up to $A \sim 1 - 2 \,\mu\mathrm{m}$ depending on specific velocity
distribution subsets and transit profiles), 
matching the sub-micrometre to micrometre scale driving amplitudes employed in q-BOUNCE setups. We have assumed
that the travelling neutrons are in oscillatory mode during all their time of flight. But the calculation can be
adapted to allow for an initial settling time,  
requiring an increase in the value of $A$ that can be derived from the previous formulae. Notice that
the optimal value of $A$ is the same for transitions between {\em any} two levels.

A further point to analyze is the mechanism by which ultra-cold neutrons entering the experimental apparatus populate
the lowest quantum levels of the gravitational potential. Neutrons emerge from delivery guide tubes in a broad,
incoherent superposition of numerous high-lying vertical states:
$  \psi(z,t)= \sum_n c_n \psi_n(z,t) $.
Because spontaneous radiative or gravitational
decay lifetimes between these micro-electronvolt states are exceedingly long (see next section),
natural relaxation is entirely negligible on the time scale of the transit. 

Instead, population of the lowest gravitational states is achieved deterministically through spatial state filtering.
The experimental apparatus utilizes a horizontal neutron absorber or rough scatterer positioned precisely a few
micrometers above the reflecting mirror, depending on the number of levels to be analyzed.
Because higher-lying quantum states possess extended spatial wavefunctions
that reach far into the vertical coordinate ($\langle n|z|n \rangle \propto \alpha_n z_0$), neutrons occupying these
excited states collide with the upper absorber and are systematically scattered out or lost from the beam. If so adjusted, 
only neutrons residing in e.g. the ground state ($n=1$), whose spatial extent is strictly confined close to the mirror surface
($z_1 \approx 9.15\,\mu\mathrm{m}$), would successfully traverse the initial collimation stage without interacting with the obstacle. 
This selective absorption window ensures that a clean, well-prepared population of the lowest gravitational
quantum state enters the resonance interaction region.

\section{The fictitious gravitational quanta}
\label{sec:grav_quanta}

It is instructive to examine how the system appears entirely from within the non-inertial frame comoving with
the oscillating boundary. In this frame, the physical mirror and its microscopic composition are absent;
instead, the system is governed by a time-dependent fictitious gravitational field arising from
the frame acceleration described by a metric perturbation $h_{\mu\nu}$.

A quantum physicist restricted entirely to this reference frame interprets transitions between the neutron's states
not as a response to a moving material boundary, but as the interaction with a time-varying background potential,
When this time-dependent potential is treated quantum-mechanically, the background can be viewed as
exchanging discrete packets of energy ---fictitious gravitons--- with the neutron. In this language:
\begin{enumerate}
\item Spatial Level Transitions ($|n\rangle \to |m\rangle$): The neutron absorbs or emits a
  gravitational quantum of energy $\hbar\omega_{nk}$ directly from the time-varying fictitious field, changing its vertical
  quantum state in the gravitational potential wells.
\item Combined Spin and Spatial Transitions:
  So far the magnetic dipole moment of the neutron has not been considered, but for later discussion it should be mentioned
  at this point.
  
  If magnetic coupling is included, the total energy change required for
  a simultaneous neutron spin flip ($\Delta E_Z = 2|\boldsymbol{\mu}_n|B_0$) and spatial state transition ($E_m - E_n$)
  is supplied by absorbing a quantum $\hbar\Omega$ from the fictitious gravitational field, while balancing energy conservation
  through the emission or absorption of a photon in the transverse electromagnetic field.
\end{enumerate}

This perspective provides an internally consistent physical picture for an observer confined to the non-inertial frame,
mirroring how the Equivalence Principle maps frame acceleration into local gravitational potentials without
requiring reference to an external laboratory mirror.

The level flip could of course be mediated by real gravitons too, but the transition probabilities
are exceedingly small. Indeed, spontaneous quadrupole transition rates mediated by thermal or
quantum graviton emission between micrometer-scale gravitational bound states scale with Newton's constant as
\begin{equation}
    \Gamma_{\text{grav}} \sim \frac{G\, m^2 \omega_{12}^5}{c^5} \langle 2| z^2 |1 \rangle^2.
\end{equation}
Substituting $m \approx 1.67 \times 10^{-27}\,\mathrm{kg}$ and $\omega_{12} \approx 1.6 \times 10^3\,\mathrm{rad/s}$
yields an estimated spontaneous emission rate of $\Gamma_{\text{grav}} \sim 10^{-76}\,\mathrm{s}^{-1}$, corresponding
to a lifetime orders of magnitude larger than the age of the Universe. 
Therefore, spontaneous decay or excitation between gravitational states via genuine gravitational
radiation is completely negligible.

\section{Vacuum radiation from a periodically oscillating boundary}
\label{sec:dce}
The previous sections establish the non-inertial description of the moving mirror, proving
that it is useful to understand the results of the q-BOUNCE experiments.
We now turn to a logically subsequent question: what happens when the electromagnetic field is quantized?

For a one-dimensional idealized moving boundary with an oscillating trajectory such
as the one in Eq. (\ref{phase}),
the creation and annihilation operators in the LAB and non-inertial frames are related by
a Bogoliubov transformation \cite{birrell,lambert}. If $a, a^\dagger$ are the operators in the LAB inertial frame
and $b, b^\dagger$ are their equivalents in the non-inertial frame they are related
by (schematically)
 $ b= \alpha a + \beta a^\dagger $.

The vacuum is defined by the condition $a \vert 0\rangle =0 $ and in the LAB frame there
are no particles. However in the non-inertial frame
$\langle 0 \vert b^\dagger b \vert 0\rangle \sim|\beta |^2$

For simplicity let us consider a massless scalar field living in the $t,z$ coordinates. Later we will
include the $x,y$ transverse coordinates. In two dimensions there is a separation into right movers and left movers.
Let us restrict the discussion to just one of them. We adopt the standard relativistic normalization of the field
modes in \(1+1\) dimensions, with the corresponding normalization of the creation and annihilation operators.
In the LAB system and non-inertial systems
\begin{equation}
  \phi(t_{\mathrm LAB},z_{\mathrm LAB})= \int_0^\infty \frac{ d\omega}{4\pi \omega}[a_\omega e^{-i \omega(t_{\mathrm LAB}-z_{\mathrm LAB}/c)}
      + a^\dagger_\omega e^{i \omega (t_{\mathrm LAB}-z_{\mathrm LAB}/c)}],
\end{equation}
\begin{equation}
\phi^\prime(t,z)= \int_0^\infty \frac{ d\omega^\prime}{4\pi \omega^\prime}[b_{\omega^\prime} e^{-i \omega^\prime(t-z/c)}
      + b^\dagger_{\omega^\prime} e^{i \omega^\prime (t-z/c)}],
  \end{equation}
respectively. Using the transformation Eqs. (\ref{phase}) and (\ref{timechange}) at lowest order in $A$
\begin{equation}
  t-z/c = t_{\mathrm LAB} -z_{\mathrm LAB}/c + \frac{A}{c} \sqrt{1 + (\frac{\Omega z_{\mathrm LAB}}{c})^2 } \sin(\Omega t_{\mathrm LAB}
    + \varphi).
  \end{equation}
where
\begin{equation}
\tan \varphi = - \frac{c}{\Omega z_{\mathrm LAB}}, \qquad R(z_{\mathrm LAB})=\frac{A}{c} \sqrt{1 + (\frac{\Omega z_{\mathrm LAB}}{c})^2 } 
  \end{equation}
Using the Jacobi-Anger expansion,
\begin{equation}
  \exp[ - i \omega^\prime R(z_{\mathrm LAB}) \sin(\Omega t_{\mathrm LAB} + \varphi(z_{\mathrm LAB}))]
    = \sum_{-\infty}^{\infty} J_n \left(\omega^\prime R(z_{\mathrm LAB})\right)\exp-in (\Omega t_{\mathrm LAB} + \varphi(z_{\mathrm LAB})). 
\end{equation}

The non-linear map mixes positive- and negative-frequency modes. The Bogoliubov coefficient
can be written in the form
\begin{equation}
\beta_{\omega^\prime,\omega}=
\frac{\omega'}{\omega}
\sum_{n\ge1}J_n\left(\omega^\prime R(z_{\mathrm LAB})\right) e^{-in\varphi(z_{\mathrm LAB})}
\delta(\omega+\omega'-n\Omega),
\end{equation}
where $J_n$ denotes a Bessel function of the first kind. 

The lowest-order harmonic therefore suggests the pair-production condition
\begin{equation}
    \omega+\omega'=\Omega,
\end{equation}
while higher harmonics obey $\omega+\omega'=n\Omega$. They are suppressed by higher powers of $A$
that are anyway beyond our approximation.

The particle production associated with this mode mixing is reminiscent of the Unruh effect in that it arises
in a non-inertial description of the quantum field. However, the mechanism considered here leads to a non-thermal
spectrum, rather than the thermal spectrum characteristic of the standard Unruh effect.

\subsection{Radiation spectrum and flux}

The previous result concerned only right-movers. Including left-movers will eventually bring a factor two
in the radiation flux. Furthermore,
the only physical particles that could be detected by the non-inertial observer are photons and gravitons
(real gravitons, that is)
but the latter have ---as we have seen--- a very small probability of interacting with the neutrons.
Consequently, the radiation in this setting consists of photons with a non-thermal spectrum. 
That would bring an additional factor two in the flux due to the two photon polarizations.

In the small amplitude regime, $J_1(x)\simeq x/2$ and we neglect all terms with $n>1$. Then the phase $\varphi(z_{\mathrm LAB})$ drops
from the probabilities and
\begin{equation}
  \beta_{\omega^\prime \omega}= \frac12\frac{(\omega^\prime)^2}{\omega}\frac{A}{c}\sqrt{1 + (\frac{\Omega z_{\mathrm LAB}}{c})^2}
  \delta(\omega + \omega^\prime -\Omega)\simeq \frac{A}{2c}\frac{(\omega^\prime)^2}{\omega} \delta(\omega +\omega^\prime -\Omega).
\end{equation}
This translates into 
\begin{equation}
  \langle 0| b^\dagger_{\omega^\prime}b_{\omega^\prime}| 0\rangle = \int d \omega \, |\beta_{\omega^\prime \omega}|^2
  \end{equation}
Let us now include the remaining degrees of freedom to compute the total flux. Integration
over the transverse degrees of freedom is also needed and provides an additional factor $(\omega^\prime)^2/ 4\pi c^2$. Finally
\begin{equation}\label{flux_spec}
  \Phi\equiv\frac{d^3 N}{d t d\omega^\prime dS}= \frac{1}{8\pi^2} \frac{(\omega^\prime)^6}{(\Omega-\omega^\prime)^2} \frac{A^2}{c^4}.
\end{equation}
Note that $\omega^\prime \le \Omega$ and that the radiation is non-thermal (unlike in the Unruh case).

This formula exhibits an apparent resonant divergence as $\omega' \to \Omega$. Strictly speaking, this divergence is an artifact
of the idealized zero-width Dirac delta function $\delta(\omega+\omega'-\Omega)$ associated with indefinitely sustained periodic motion.

The frequency integral of the emitted energy will likewise be dominated by the contribution
near the upper-frequency endpoint. The result will be sensitive to the finite duration of the
oscillatory motion, which regularizes the idealized spectrum obtained for an indefinitely sustained
periodic motion. To estimate this effect, we introduce a frequency interval \(\Delta\omega'=2\pi/\tau\),
where \(\tau\) represents the characteristic duration over which the neutron can undergo
the assumed oscillatory motion. This choice is motivated by the energy-time uncertainty relation
and provides an estimate of the frequency resolution associated with a finite observation time.
We emphasize that this prescription is intended as a physically motivated estimate of the
finite-time effect, rather than a unique determination of the spectral linewidth.

In order to maximize the energy flux and energy density one could contemplate using a trap to keep the
neutrons in place rather than allowing them to fly free. In the following we will use $\tau=100$ s in this situation.

\subsection{Vacuum energy and magnetic fields}

Let us evaluate the magnitude of the Unruh-like electromagnetic field experienced directly by the accelerated neutron.
Starting from the spectral number flux per unit area in Eq.~(\ref{flux_spec}), the total radiated energy flux
  (power per unit area) integrated up to the regularized upper cutoff $\omega'_{\max} = \Omega - \Delta\omega'$, with $\Delta\omega' = 2\pi/\tau$, is
\begin{equation}
    \mathcal{F}_E = \int_0^{\Omega - \Delta\omega'} d\omega'\, \hbar \omega' \Phi(\omega')
    = \frac{\hbar A^2 \Omega^6}{8\pi^2 c^4} \int_0^{1-\delta} dx\, \frac{x^7}{(1-x)^2},
\end{equation}
where $x = \omega'/\Omega$ and $\delta \equiv \Delta\omega'/\Omega = \frac{2\pi}{\Omega\tau} \ll 1$.
In the long-interaction regime $\Omega\tau \gg 1$, the integral is dominated by the resonant pole $1/\delta$, yielding
\begin{equation}
    \mathcal{F}_E \simeq \frac{\hbar A^2 \Omega^7 \tau}{16\pi^3 c^4}
    \left[ 1 - \frac{14\pi}{\Omega\tau} \ln\!\left(\frac{\Omega\tau}{2\pi}\right) + \mathcal{O}\!\left(\frac{1}{\Omega\tau}\right) \right].
\end{equation}
The log and subsequent terms are clearly subdominant.
Dividing by the speed of light gives the volumetric electromagnetic energy density $u = \mathcal{F}_E / c$ of the vacuum
radiation bath in the neutron's non-inertial frame:
\begin{equation}\label{energy_density}
    u \simeq \frac{\hbar A^2 \Omega^7 \tau}{16\pi^3 c^5}.
\end{equation}
Since electromagnetic energy density is shared equally between electric and magnetic components ($u = u_E + u_B = B_{\mathrm{rms}}^2/\mu_0$),
the root-mean-square magnetic field fluctuation felt by the neutron in free space is
\begin{equation}\label{B_rms_free}
    B_{\mathrm{rms}} = \sqrt{\mu_0 u} \simeq \frac{A}{4\pi} \sqrt{\frac{\mu_0 \hbar \Omega^7 \tau}{\pi c^5}},
\end{equation}
where $\mu_0$ is the magnetic permeability.
Although each produced photon carries a magnetic field oriented randomly in the transverse $xy$-plane, $B_{\mathrm{rms}}$ sets
the characteristic field amplitude. This $B_{\mathrm{rms}}$ could eventually couple to the neutron's own
magnetic moment and possibly induce spin flips.

We will see later the order of magnitude predicted for $B_{\mathrm{rms}}$.

\section{Coupling to the neutron magnetic moment}
\label{sec:spin}

Although the neutron is electrically neutral, it possesses an intrinsic magnetic dipole moment,
\begin{equation}
    \boldsymbol{\mu}_n
    =g_n\mu_N\frac{\mathbf S}{\hbar},
    \qquad
    \mu_N=\frac{e\hbar}{2m_p},
    \qquad
    g_n\simeq-3.826.
\end{equation}
The interaction with an oscillating magnetic field is governed by
\begin{equation}\label{withB}
    H_B(t)=-\boldsymbol{\mu}_n\cdot\mathbf B(t).
\end{equation}

When the neutron is immersed in a weak static holding field $\mathbf{B}_0$
in the $\mathbf z$ direction defining its spin quantization axis, each gravitational
eigenstate $|n\rangle$ of Sec.~\ref{sec:boundstates} splits into a Zeeman doublet $|n, \uparrow\rangle$ and $|n, \downarrow\rangle$ with energies
\begin{equation}
    E_{n,\pm} = E_n \pm |\boldsymbol{\mu}_n| B_0 = \hbar\!\left(\omega_0 \alpha_n \pm \frac{\omega_L}{2}\right),
\end{equation}
where $\omega_L = 2|\boldsymbol{\mu}_n|B_0/\hbar$ is the Larmor frequency. Between any two gravitational levels $n$ and $k$,
the Zeeman splitting lifts the spin degeneracy into three distinct transition frequencies: the spin-conserving doublet
$\Omega_0 = \omega_{nk}$ ($\Delta s = 0$) and two spin-flip sidebands $\Omega_{\pm} = \omega_{nk} \pm \omega_L$ ($\Delta s = \pm 1$).

The field $h_{00}$ is spin-independent and it 
excites the $\Delta s = 0$ transitions only.
Substituting the general dipole matrix element Eq.~(\ref{airy_dipole}) at $\Omega = \omega_{nk}$ gives the resonant Rabi frequency
derived in Section II. The optimal value for the amplitude $A$ to drive the population of the first two levels has been derived
also there for a given transit time of the neutrons across the experimental set-up.

To access the entire spin--gravitational Hilbert space $|n, s\rangle$  it suffices to superpose a spatially uniform oscillating
transverse radio-frequency (RF) field in the horizontal $xy$-plane,
$\mathbf{B}_{\mathrm{RF}}(t) = B_1 \cos(\omega_{\mathrm{RF}} t_{\mathrm LAB})\hat{\mathbf{x}}$.

In the comoving non-inertial frame, the neutron's magnetic dipole interacts with the electromagnetic field through the covariant
Pauli coupling, $-\frac{\mu_n}{2}\sqrt{-g}\,\bar{\psi}\sigma^{\mu\nu}\psi\,F_{\mu\nu}$, in the curved background metric of Sec.~\ref{sec:noninertial}.
Transforming the electromagnetic field tensor under Eqs.~(\ref{phase}) and (\ref{timechange}), and expanding the metric determinant
and tetrads in powers of the oscillation amplitude $A$, allows us to identify the leading interaction and the relativistic corrections.

At zeroth order in $A$, the transverse RF magnetic field remains spatially uniform and does not couple to the vertical motion.
The corresponding Hamiltonian is
\begin{equation}
    H_{\mathrm{RF}}(t) = |\boldsymbol{\mu}_n| B_1 \cos(\omega_{\mathrm{RF}} t)\hat{\sigma}_x,
\end{equation}
which acts exclusively in spin space ($\langle k | n \rangle = \delta_{nk}$) and drives pure spin-flip transitions
($|n,\downarrow\rangle \leftrightarrow |n,\uparrow\rangle$, $\Delta n = 0$, $\Delta s = \pm 1$) at $\omega_{\mathrm{RF}} = \omega_L$,
with Rabi frequency $\Omega_{\mathrm{spin}} = |\boldsymbol{\mu}_n| B_1 / \hbar$.

At first order in $A$, two relativistic effects modify the RF interaction. First, the vertical frame velocity $v_z(t) = -A\Omega\sin(\Omega t)$
induces a transverse electric field $E'_y \simeq v_z B_x$ in the comoving frame, but since the neutron is electrically neutral, this field
does not produce any relevant effect. The RF magnetic field itself is unchanged at this order, $B'_x = B_x + \mathcal{O}(A^2)$.
Second, the clock-synchronization shift, $t_{\mathrm{LAB}} \simeq t - (A\Omega z / c^2)\sin(\Omega t)$, introduces a position-dependent
modulation of the RF phase. The resulting coupling between the vertical coordinate and the spin is of
order $A\Omega\omega_{\mathrm{RF}}z_0/c^2 \sim 10^{-22}$ and is therefore negligible.

The non-inertial metric also produces a position-dependent correction to the magnetic dipole interaction through $h_{00}$,
corresponding to the gravitational redshift of the dipole energy. Expanding the covariant interaction to first order in $h_{00}$ gives
\begin{equation}
\frac{1}{2}\int d^3x\,T_{\mathrm{dip}}^{00}h_{00}
= (\boldsymbol{\mu}_n \cdot \mathbf{B})
\frac{A\Omega^2 z}{c^2}\cos(\Omega t).
\end{equation}
This term couples the spin to the vertical motion, but its magnitude is likewise negligible, at the $\sim 10^{-22}$ level for the
parameters considered. 
As for the spatial-stress contribution $\frac{1}{2}\int d^3x\,T_{\mathrm{dip}}^{zz}h_{zz}$ is second order in the oscillation
amplitude and is neglected consistently with our first-order approximation.

Thus, the exact non-inertial Hamiltonian does not strictly factorize into independent spin and gravitational parts: relativistic
corrections introduce weak couplings between the two degrees of freedom, and hence a small spin--gravitational entanglement.
This is the Equivalence-Principle counterpart of the frame-dependent simultaneity effects in the inertial LAB description.

In practice, however, these corrections are utterly negligible, and the spin and gravitational dynamics can be treated as independent.
For all practical purposes we can consider that $H_{\mathrm int}(t)$ defined in Sect III and the magnetic hamiltonian,
approximated by
\begin{equation}
    H_{\mathrm{RF}}(t) = |\boldsymbol{\mu}_n| B_1 \cos(\omega_{\mathrm{RF}} t)\hat{\sigma}_x,
\end{equation}
commute and $H_{\mathrm RF}$ acts exclusively in spin space ($\langle k | n \rangle = \delta_{nk}$) and driving
pure spin-flip transitions
($|n,\downarrow\rangle \leftrightarrow |n,\uparrow\rangle$, $\Delta n = 0$, $\Delta s = \pm 1$) at $\omega_{\mathrm{RF}} = \omega_L$
with Rabi frequency $\Omega_R^{\mathrm{spin}} = |\boldsymbol{\mu}_n| B_1 / \hbar$.

Because $H_{\mathrm int}(t)$ and the uniform RF drive $H_{\mathrm{RF}}(t)$ act on decoupled Hilbert spaces and commute at
all times, the combined time-evolution operator factorizes into independent spatial and spin rotations.
Applying both drives simultaneously at their respective resonances ($\Omega = \omega_{nk}$ and $\omega_{\mathrm{RF}} = \omega_L$)
yields the joint transition probability
\begin{equation}
  P_{|n,\downarrow\rangle \to |k,\uparrow\rangle}(\tau) = \
  \sin^2\!\left(\frac{\Omega_R \tau}{2}\right)\sin^2\!\left(\frac{\Omega_{\mathrm{spin}} \tau}{2}\right).
\end{equation}
Full population transfer to any target state $|k, s'\rangle$ is thus achieved simply by maintaining the optimal
mechanical amplitude $A_{\mathrm{opt}} = \pi g / (2\omega_0^3 \tau)$ of Sec.~\ref{sec:transition} while setting the uniform RF field
amplitude to $B_1 = \pi\hbar / (|\boldsymbol{\mu}_n|\tau)$.
By contrast, exciting the combined transitions $|n, s\rangle \to |k, s'\rangle$ directly at the Zeeman-shifted sideband frequencies
$\Omega_{\pm} = \omega_{nk} \pm \omega_L$ using a single monochromatic field—rather than two separate drives at $\omega_{nk}$ and
$\omega_L$—requires breaking this factorization via a non-zero spatial gradient $\partial B_x/\partial z$ (producing an interaction
proportional to $z \otimes \hat{\sigma}_x$).

Beyond serving as a negligible correction to the driven Rabi dynamics, the gravitational-redshift coupling
$\frac{1}{2}\int d^3x\,T_{\mathrm{dip}}^{00}h_{00}$ carries a broader conceptual significance for Quantum Equivalence Principle (QEP)
tests using quantized gravitational states. Because the neutron's internal Zeeman energy $E_{\mathrm{spin}} = \mp\hbar\omega_L/2$
contributes to its effective gravitational and inertial mass ($\Delta m = E_{\mathrm{spin}}/c^2$), the spin-up and spin-down components
experience slightly different effective accelerations $a_{\uparrow,\downarrow}(t) = a(t)\,[1 \mp \hbar\omega_L/(2mc^2)]$ as well as
a static Earth gravitational redshift $\Delta\omega_L/\omega_L = g(\langle z\rangle_k - \langle z\rangle_n)/c^2$ across distinct
gravitational eigenstates $|n\rangle$ and $|k\rangle$.

Consider an initial state consisting of a superposition of two spin states, \(|\Psi(0)\rangle=(|n,+\rangle+|n,-\rangle)/\sqrt{2}\),
where the gravitational state can initially be factorized from the spin state. The spin-dependent gravitational interaction may, however,
induce transitions to other gravitational states, with amplitudes that depend on the spin component. The resulting state can then become
entangled, such that the spin and gravitational degrees of freedom can no longer be described independently.
The degree of entanglement is determined by how differently the gravitational state evolves for the two spin components.
For the parameters considered here,
these effects are expected to be exceedingly small.

\subsection{The effect of $B_{\mathrm{rms}}$}

As shown in the previous section, transversal magnetic fields are generated by a mechanism analogous, but distinct, to
the Unruh effect. These magnetic fields are felt by the neutron only, which is the only object that is at rest in
the non-inertial frame. The magnetic fields are distributed over a range of frequencies, but peaked at $\omega=\Omega$,
the externally driven frequency.

Let us now revert to the simplest where the transverse field is set to zero $B_1=0$ and we are
interested in the possibility that the transverse RMS magnetic field observed in 
the non-inertial reference frame is able to trigger a spin flip.

When the oscillation frequency matches the Larmor resonance associated
to the holding field $B_0$ ($\Omega \simeq \omega_L$), the spin-flip transition probability grows
coherently over an interaction time $\tau$ as
\begin{equation}
    P_{\uparrow\to\downarrow}(\tau) = \sin^2\left(\frac{\bar\Omega_R \tau}{2}\right) \simeq \left(\frac{\bar\Omega_R \tau}{2}\right)^2,
    \qquad
    \bar\Omega_R = \frac{|g_n| \mu_N B_{\mathrm{rms}}}{\hbar},
\end{equation}
provided $\bar\Omega_R \tau \ll 1$. It is important to distinguish this spin-flip channel from the gravitational-state
transitions of Sec.~\ref{sec:transition}: the former probes the transverse magnetic fluctuations of the Unruh-like
radiation, whereas the latter is driven by the longitudinal fictitious gravitational potential.

We can now evaluate Eq.~(\ref{B_rms_free}) for a mechanically driven q-BOUNCE setup. Taking the $|1\rangle \to |3\rangle$
mechanical drive at the optimal $\pi$-pulse amplitude of Sec.~\ref{sec:transition} as an example,
\begin{equation}
    \frac{\Omega}{2\pi}\simeq 462.7\,\mathrm{Hz} \quad (\Omega \simeq 2907\,\mathrm{rad/s}),
    \qquad A = 772\,\mathrm{nm}, \qquad
    a_{\max} = A\Omega^2 \simeq 6.5\,\mathrm{m/s^2},
\end{equation}
and assuming an idealized storage duration of $\tau = 100\,\mathrm{s}$, the energy density and magnetic field evaluate to
\begin{equation}
    u \simeq 9.2 \times 10^{-66}\,\mathrm{J/m^3},
    \qquad
    B_{\mathrm{rms}} \simeq 3.4 \times 10^{-36}\,\mathrm{T}
\end{equation}
(and even smaller, $B_{\mathrm{rms}} \sim  10^{-38}\,\mathrm{T}$, over the single-pass transit time $\tau \simeq 0.026\,\mathrm{s}$).

\section{Magnetic acceleration and the gedankenexperiment threshold}
\label{sec:magnetic}

Because mechanical actuators are bounded by stress limits to accelerations of order $a_{\max} \sim 10^2\,\mathrm{m/s^2}$,
achieving substantially larger accelerations requires replacing mechanical motion with inhomogeneous magnetic fields,
a technique pioneered in UCN spectroscopy by the GRANIT collaboration \cite{GRANIT}. While our radiation analysis is
formulated in the neutron's non-inertial rest frame, the laboratory magnetic profile need only be specified here to
determine the neutron's trajectory $z_{\mathrm{LAB}}(t_{\mathrm{LAB}}) = A\cos(\Omega t_{\mathrm{LAB}})$ and peak acceleration
$a_{\max}$, which then enter directly into the comoving vacuum-radiation expressions of Eqs.~(\ref{energy_density}) and (\ref{B_rms_free}).

To preserve a well-defined spin quantization axis and avoid non-adiabatic depolarization at magnetic field
zero-crossings, we consider an orthogonal field configuration consisting of a weak static holding field $B_0$ in the
transverse plane and a strong oscillating vertical gradient $G_0 = \partial B_z / \partial z$ along $\hat{\mathbf{z}}$:
\begin{equation}
    \mathbf{B}(z_{\mathrm{LAB}}, t_{\mathrm{LAB}}) = B_0\,\hat{\mathbf{x}} + G_0 z_{\mathrm{LAB}} \cos(\Omega t_{\mathrm{LAB}})\,\hat{\mathbf{z}}.
\end{equation}
The field configuration above is understood as an idealized local description of the magnetic gradient; the additional field components
required by Maxwell's equations in a realizable magnet are not explicitly included here. 
For the purpose of estimating the attainable acceleration, we assume an effective spin projection along the gradient direction and
use the corresponding Stern–Gerlach force to prescribe the neutron trajectory. A fully self-consistent treatment of the spin dynamics
in the combined holding and oscillating fields is beyond the scope of this estimate.

Because $|\mathbf{B}| = \sqrt{B_0^2 + (G_0 z_{\mathrm{LAB}} \cos\Omega t_{\mathrm{LAB}})^2} \ge B_0 > 0$ at all times, the
total magnetic field never vanishes even as its vertical gradient reverses sign every half-cycle, while $B_0$ can be kept
small enough that the Larmor frequency $\omega_L = 2|\boldsymbol{\mu}_n|B_0/\hbar$ remains matched to the modulation frequency $\Omega$.
The weak holding field provides a reference direction for the average spin polarization, which points along $+\hat{\mathbf{x}}$ for $B_0>0$,
whereas the strong oscillating gradient is responsible for driving the vertical motion of the neutron at frequency $\Omega$. This
configuration allows us to consider accelerations much larger than those achieved in the mechanical bouncing setup and thereby estimate
the corresponding $B_{\rm rms}$. Although a sufficiently large $B_{\rm rms}$ could in principle perturb the spin polarization
relative to the holding-field direction, the value obtained below is exceedingly small, making such a perturbation mostly
of academic interest.

For spin-polarized neutrons, the vertical Stern--Gerlach force $F_z(t) = -|\boldsymbol{\mu}_n| G_0 \cos(\Omega t)$ drives a
harmonic vertical displacement $z_{\mathrm{LAB}}(t) = A\cos(\Omega t_{\mathrm{LAB}})$ with amplitude
\begin{equation}
    A = \frac{|\boldsymbol{\mu}_n| G_0}{m \Omega^2},
    \qquad
    a_{\max} = A\Omega^2 = \frac{|\boldsymbol{\mu}_n|}{m} G_0,
\end{equation}
where for the neutron $|\boldsymbol{\mu}_n|/m \simeq 5.77\,\mathrm{m^2\,s^{-2}\,T^{-1}}$. Substituting these trajectory parameters
into Eqs.~(\ref{energy_density}) and (\ref{B_rms_free}) directly yields the free-space Unruh-like energy density and RMS magnetic
field in the neutron's comoving frame:
\begin{equation}\label{B_rms_G0}
    u \simeq \frac{\hbar}{16\pi^3 c^5} \left(\frac{|\boldsymbol{\mu}_n|}{m}\right)^2 G_0^2 \Omega^3 \tau,
    \qquad
    B_{\mathrm{rms}} \simeq \frac{1}{4\pi} \left(\frac{|\boldsymbol{\mu}_n|}{m}\right) G_0 \sqrt{\frac{\mu_0 \hbar \Omega^3 \tau}{\pi c^5}}.
\end{equation}

State-of-the-art micro-fabricated magnetic structures could achieve local field gradients up to $G_0 \sim 10^4 - 5 \times 10^4\,\mathrm{T/m}$ \cite{maxG0},
corresponding to peak accelerations $a_{\max} \sim 5.8 \times 10^4 - 2.9 \times 10^5\,\mathrm{m/s^2}$. Choosing an oscillation amplitude
$A = 1\,\mu\mathrm{m}$, the maximum gradient $G_0 = 5 \times 10^4\,\mathrm{T/m}$ gives a modulation frequency
$\Omega \approx 5.4 \times 10^5\,\mathrm{rad/s}$ ($85\,\mathrm{kHz}$). The corresponding resonant holding field,
determined by $\omega_L = \Omega$, is $B_0 \approx 2.9\,\mathrm{mT}$, whereas the peak oscillating gradient
field is $G_0 A = 50\,\mathrm{mT}$. Thus the holding field is substantially weaker than the oscillating field, while ensuring that
the total magnetic field never vanishes. Assuming that neutrons can be captured for a storage time $\tau = 100\,\mathrm{s}$, Eq.~(\ref{B_rms_G0}) gives
\begin{equation}
    B_{\mathrm{rms}} \simeq 3.7 \times 10^{-28}\,\mathrm{T}.
\end{equation}
Although magnetic driving enhances the Unruh-like radiation field by about eight orders of magnitude over mechanical mirrors,
the resulting open-space field remains approximately twenty-three orders of magnitude below the terrestrial magnetic background.

\subsection{Gedankenexperiment: Threshold for an observable radiation field}

To establish the physical boundary between unobservable laboratory regimes and genuine Unruh phenomenology, we invert
Eq.~(\ref{B_rms_free}) as a theoretical benchmark: what oscillation parameters are required for an accelerated neutron
to experience a macroscopic, millitesla-scale ($B_{\mathrm{rms}} = 1\,\mathrm{mT}$) Unruh-like magnetic field in open space over $\tau = 100\,\mathrm{s}$?

Solving Eq.~(\ref{B_rms_free}) for the angular frequency at fixed amplitude $A = 100\,\mathrm{nm}$ yields
\begin{equation}
    \Omega = \left( \frac{16\pi^3 c^5 B_{\mathrm{rms}}^2}{\mu_0 \hbar A^2 \tau} \right)^{1/7} \simeq 9.86 \times 10^{12}\,\mathrm{rad/s}
    \qquad \left(f = \frac{\Omega}{2\pi} \simeq 1.57\,\mathrm{THz}\right).
\end{equation}
At this terahertz modulation frequency, the peak velocity remains non-relativistic ($v_{\max} = A\Omega \simeq 9.9 \times 10^5\,\mathrm{m/s} \approx 0.0033\,c$,
preserving the validity of our linear-in-$A$ metric expansion), while the peak acceleration reaches
\begin{equation}
    a_{\max} = A\Omega^2 \simeq 9.7 \times 10^{18}\,\mathrm{m/s^2}.
\end{equation}
Achieving this acceleration via Stern--Gerlach magnetic driving would require an extreme field gradient of
\begin{equation}
    G_0 = \frac{a_{\max}}{|\boldsymbol{\mu}_n|/m} \simeq 1.7 \times 10^{18}\,\mathrm{T/m},
\end{equation}
roughly fourteen orders of magnitude beyond current laboratory micro-magnet capabilities.

This hierarchy of scales highlights a fundamental contrast between neutral and charged probes of the Unruh effect.
For electrons, the large charge-to-mass ratio ($e/m_e \approx 1.76 \times 10^{11}\,\mathrm{C/kg}$) allows ultra-intense
laser wakefields ($E \sim 10^{11}\,\mathrm{V/m}$) to reach $a \sim 10^{21}\,\mathrm{m/s^2}$, bringing Unruh-like radiation
within experimental reach at the cost of separating a small quantum signal from overwhelming classical Larmor
radiation \cite{chen_tajima}. For electrically neutral ultra-cold neutrons, classical charge-radiation backgrounds vanish entirely,
providing a conceptually pure realization of the Equivalence Principle; however, because the magnetic dipole
coupling $|\boldsymbol{\mu}_n|/m_n$ is ten orders of magnitude smaller than $e/m_e$ and couples to field gradients rather than uniform fields,
direct detection of Unruh-like radiation in UCN setups remains squarely in the realm of a \textit{gedankenexperiment}.

\subsection{Experimental status of the Unruh effect}
To place these \textit{gedankenexperiment} bounds in broader context, it is useful to summarize the current experimental status of the
Unruh effect \cite{crispino_rmp}. Because the Unruh temperature $T_U = \hbar a / (2\pi c k_B)$ scales with only $4 \times 10^{-21}\,\mathrm{K\cdot s^2/m}$,
direct and unambiguous observation of genuine vacuum thermality in flat spacetime remains an open experimental challenge.
Existing and proposed efforts fall into three main categories:

\begin{enumerate}

\item Analog gravity platforms: Systems such as Bose--Einstein condensates (BECs), superconducting circuits, and nonlinear optical
  fibers reduce the effective speed of signal propagation, lowering the acceleration threshold required for thermalization.
  While such experiments have successfully observed analog Hawking and Unruh-like phononic/photonic spectra \cite{hu_bec},
  they test kinematic quantum field theory in effective media rather than the fundamental electromagnetic vacuum coupled to
  real spacetime acceleration.

\item High-energy charged particles: The most extreme proper accelerations accessible in the laboratory occur for
  relativistic electrons or positrons undergoing violent radiation recoil in aligned crystals (such as the CERN-NA63 channeling experiments)
  \cite{lynch_unruh} or ultra-intense laser fields \cite{chen_tajima}. Thermalized photon spectra consistent with Fulling--Davies--Unruh
  temperatures have been reported in high-energy channeling radiation \cite{lynch_unruh}, though interpreting these signatures as direct proof
  of the Unruh effect rather than standard quantum electrodynamics in strong external fields remains a subject of active
  theoretical discussion \cite{crispino_rmp} to this day.

\item Low-energy neutral quantum probes: Cold atoms and ultra-cold neutrons eliminate Coulomb interactions and classical
  Larmor radiation backgrounds entirely, offering the cleanest conceptual test of the Equivalence Principle. However, as our
  estimates demonstrate, the smallness of the magnetic dipole coupling ($|\boldsymbol{\mu}_n|/m_n$) and laboratory limits on magnetic
  field gradients confine direct detection of Unruh-like radiation in UCN setups to the \textit{gedankenexperiment} regime.

\end{enumerate}

\section{Conclusions}

In this work, we have revisited the quantum dynamics of ultra-cold neutrons in gravitational bound states from the
foundational perspective of the General Relativity Equivalence Principle. By transforming into the non-inertial
reference frame comoving with a periodically vibrating boundary, we have established a 
field-theoretic description where the boundary motion is equivalent to a time-dependent metric
perturbation $h_{\mu\nu}$ and a fictitious gravitational potential.

Following this logic, we have examined the experimental setting entirely from
the point of view of the non-inertial frame. While standard q-BOUNCE experiments rely on mechanical
piezoelectric mirrors or magnetic devices to drive transitions between gravitational Airy states, applying quantum
field theory in the accelerated frame predicts an accompanying Unruh-like bath of electromagnetic vacuum radiation
that couples to the neutron's magnetic dipole moment.

Let us summarize our main findings:
\begin{enumerate}
\item \textit{Consistency of non-inertial descriptions:} The frame transformation cleanly accounts for transitions
  between the neutron Airy states without requiring explicit reference to an external laboratory mirror, reinterpreting level
  jumps as the exchange of discrete fictitious gravitational quanta. Key experimental parameters—including resonant
  driving frequencies, optimal sub-micrometre oscillation amplitudes (including Zeeman splitting in a uniform holding and
  transverse RF field) —are naturally understood in this picture.
    
\item \textit{Unruh-like radiation from mechanical boundaries:} Neutrons whose vertical wave function oscillates experience
  an electromagnetic radiation flux in their non-inertial frame. However, standard mechanically driven mirrors, bounded by
  piezoelectric acceleration limits ($a_{\max} \sim 200\,\mathrm{m/s^2}$), generate an open-space magnetic field of only
  $B_{\mathrm{rms}} \sim 10^{-36}\,\mathrm{T}$, far too weak to induce observable spin-flip transitions.
    
\item \textit{Viability of magnetic parametric driving:} Replacing mechanical boundary motion with high-gradient
  periodic magnetic fields ($G_0 \sim 10^4\,\mathrm{T/m}$) orthogonal to a weak static holding field
  bypasses mechanical stress limits, increasing effective accelerations by orders of magnitude
  and enhancing the open-space Unruh-like magnetic field to $B_{\mathrm{rms}} \sim 10^{-28} - 10^{-27}\,\mathrm{T}$. Still
  many orders of magnitude too small.
    
\item \textit{Gedankenexperiment threshold for observability:} Despite this enhancement, open-space Unruh-like radiation
  in realistic UCN setups remains many orders of magnitude below terrestrial and shielded magnetic backgrounds. Inverting the
  radiation flux as a \textit{gedankenexperiment} shows that generating a macroscopic millitesla-scale radiation field in free
  space requires terahertz modulations ($\Omega \sim 10^{13}\,\mathrm{rad/s}$), accelerations of order $a \sim 10^{19}\,\mathrm{m/s^2}$,
  and in the case of magnetic driving with magnetic gradients $G_0 \sim 10^{18}\,\mathrm{T/m}$.
\end{enumerate}

These results demonstrate that while UCN gravitational traps remain an ideal laboratory for probing gravitational
level quantization, testing the Weak Equivalence Principle, and constraining short-range forces, the weak magnetic dipole
coupling of neutral neutrons confines direct detection of Unruh-like radiation to the \textit{gedankenexperiment} regime—in
sharp contrast to charged electrons in ultra-intense laser wakefields, where extreme proper accelerations are achievable.

\section*{Acknowledgements}
 The financial support from
the State Agency for Research of the Spanish Ministry of Science, Innovation and Universities through the “Unit of Excellence Maria de Maeztu”
award to the Institute of Cosmos Sciences (CEX2024-001451-M) and through project PID2022-136224NB-C21 is acknowledged. We also acknowledge
the support of the Catalan Government through grant 2021-SGR-249.
The author thanks the CERN TH Division, where the final version
of this work took form, for hospitality.


\begin{thebibliography}{9}

\bibitem{nature} V. V. Nesvizhevsky et al., Nature \textbf{415}, 297 (2002); 
V. V. Nesvizhevsky et al., Phys. Rev. D \textbf{67}, 102002 (2003);
V. V. Nesvizhevsky et al., Eur. Phys. J. C \textbf{40}, 479 (2005);
 T. Jenke et al., EPJ Web of Conferences \textbf{219}, 05003 (2019).

\bibitem{pioneer} V. I. Lushchikov and A. I. Frank, JETP Lett. \textbf{28}, 559 (1978).

\bibitem{details} A. Westphal et al. Eur. Phys. J. C \textbf{51}, 367 (2007)

\bibitem{name} H. Abele et al. TU Wien repositum (Report INIS-AT 2002-0004);
  T. Jenke, D. Stadler, H. Abele and P. Geltenbort, Nucl. Instr. and Methods Sect A \textbf{611}, 318 (2009).

\bibitem{WEP}  
O. Bertolami and F. M. Nunes, Class. Quantum Grav. \textbf{20}, L61 (2003);
  
\bibitem{5th}
H. Abele, T. Jenke, D. Stadler and P. Geltenbort, Nucl. Phys. A \textbf{827}, 593 (2009)
 H. Abele, T. Jenke, H. Leeb, and J. Schmiedmayer,
 Phys. Rev. D \textbf{81}, 065019 (2010);
 G. Cronenberg et al., Nature Phys. \textbf{14}, 1022--1026 (2018).

\bibitem{unruh} W. G. Unruh, Phys. Rev. D \textbf{14}, 870 (1976).

\bibitem{moore} G. T. Moore, J. Math. Phys. \textbf{11}, 2679 (1970).
  
\bibitem{DCE} V. V. Dodonov, Phys. Scr. \textbf{82}, 038105 (2010).

  \bibitem{chalmers} C. M. Wilson et al., Nature \textbf{479}, 376 (2011).

\bibitem{ignatovich} V. K. Ignatovich, \textit{The Physics of Ultracold Neutrons} (Clarendon Press, Oxford, 1990).

\bibitem{birrell} S. A. Fulling and P. C. W. Davies,
Proc. R. Soc. London, Ser. A \textbf{348}, 393 (1976);
N. D. Birrell and P. C. W. Davies, Quantum Fields in Curved Space (Cambridge University Press, 1982).

\bibitem{lambert} A. Lambrecht, M. T. Jaekel and S.Reynaud, Phys. Rev Lett. 77, 615 (1996).
  

\bibitem{GRANIT}
M. Kreuz et al., Nucl. Instrum. Methods Phys. Res. A \textbf{611}, 326 (2009);
V. V. Nesvizhevsky et al., Nature Phys. \textbf{6}, 114 (2010);
S. Baessler et al., C. R. Physique \textbf{12}, 729 (2011);
G. Pignol et al., Adv. High Energy Phys. \textbf{2014}, 628125 (2014);
D. Roulier et al., Adv. High Energy Phys. \textbf{2015}, 730437 (2015).
  

\bibitem{maxG0} L. Schein-Lubomirsky, Y. Mazor, R. Stöhr et al., Commun. Phys. \textbf{8}, 98 (2025)
  A. Sathyadharma Prasad, R. Ravishankar, B. Büchner et al., Commun. Mater. \textbf{6}, 164 (2025)

\bibitem{crispino_rmp} L.~C.~B. Crispino, A. Higuchi, and G.~E.~A. Matsas, Rev. Mod. Phys. \textbf{80}, 787 (2008).
  
\bibitem{hu_bec} J. Hu, L. Feng, Z. Zhang, and C. Chin, Nat. Phys. \textbf{15}, 785 (2019).
  
\bibitem{lynch_unruh} M.~H. Lynch, E. Cohen, Y. Hadad, and I. Kaminer, Phys. Rev. D \textbf{104}, 025015 (2021).

\bibitem{chen_tajima} P. Chen and T. Tajima, Phys. Rev. Lett. \textbf{83}, 256 (1999).



\end{thebibliography}
\end{document}